\documentclass[12pt]{article}
\usepackage[english]{babel}
\usepackage{mathtext}
\usepackage[T2A]{fontenc}
\usepackage{epsfig}
\begin{document}
\begin{center}
{\Large {\bf  Renyi Thermostatistics and Self-Organization}}\\
\renewcommand{\thefootnote}{\fnsymbol{footnote}}
\vspace{.5cm} \large {A.G.Bashkirov}\footnote{{\it E-mail
address}: abas@idg.chph.ras.ru}\\ \vspace{.25cm} Institute
Dynamics of Geospheres, RAS,\\ Leninskii prosp. 38 (bldg.1),
119334, Moscow, Russia\\
\end{center}

\begin{abstract}

Taking into account extremum of a Helmholtz free energy in the
equilibrium state of a thermodynamic system the Renyi entropy is
derived from the Boltzmann entropy by the same way as the
Helmholtz free energy from the Hamiltonian. The application of
maximum entropy principle to the Renyi entropy gives rise to the
Renyi distribution. The $q$-dependent Renyi thermodynamic entropy
is defined as the Renyi entropy for Renyi distribution. A
temperature and free energy are got for a Renyi thermostatistics.
Transfer from the Gibbs to Renyi thermostatistics is found to be a
phase transition at zero value of an order parameter $\eta=1-q$.
It is shown that at least for a particular case of the power-law
Hamiltonian $H=C\sum_i x_i^\kappa$ this entropy increases with
$\eta$. Therefore in the new entropic phase at $\eta>0$ the system
tends to develop into the most ordered state at
$\eta=\eta_{max}=\kappa/(1+\kappa)$. The Renyi distribution at
$\eta_{max}$ becomes a pure power-law distribution.
\end{abstract}


\section{Introduction}

Numerous examples of power--law distributions (PLD) are well-known
in different fields of science and human activity \cite{Mant}.
Power laws are considered \cite{Bak} as one of signatures of
complex self-organizing systems. They are sometimes called
Zipf--Pareto law or fractal distributions. We can mention here the
Zipf--Pareto law in linguistics \cite{Zipf}, economy \cite{Mand2}
and  in the science of sciences \cite{Price}, Gutenberg-Richter
law in geophysics \cite{Gut}, PLD in critical phenomena
\cite{Baxt}, PLD of avalanche sizes in sandpile model for
granulated media \cite{Bak1} and fragment masses in the impact
fragmentation \cite{Fujiw,fragm}, etc.

According to the well-known maximum entropy principle  developed
by Jaynes \cite{Jaynes} for a Boltzmann-Gibbs statistics
an equilibrium distribution of probabilities 
must
provide maximum of the Boltzmann  information entropy $S_B$ upon
additional conditions of normalization  $\sum_i p_i=1$ and a fixed
average energy $ U=\langle H\rangle_p\equiv\sum_i H_i p _i$.

Then, the Gibbs canonical distribution $\{p^{(G)}_i\}$ is
determined from the extremum of the functional
 \begin{equation}
L_G(p )=- \sum_i^W\,p_i\ln p_i - \alpha_0 \sum_i^W\,p _i - \beta_0
\,\sum_i^W H_i p _i ,
\end{equation}
where $ \alpha_0 $ and $\beta_0 =1/k_BT_0 $  are Lagrange
multipliers and $T_0$ is the thermodynamic temperature.

However, when investigating complex physical systems (for example,
fractal and self-organizing structures, turbulence) and a variety
of social and biological systems, it appears that the Gibbs
distribution does not correspond to observable phenomena. In
particular, it is not compatible with a power-law distribution
that is typical \cite{Bak}  for such systems. Introducing of
additional constraints on a sought distribution in the form of
conditions of true average values $\langle X^{(m)}\rangle_p$ of
some physical parameters of the system $X^{(m)}$ gives rise to a
generalized Gibbs distribution with additional terms in the
exponent but does not change its exponential form.

Montroll and Shlesinger \cite{Montr} investigated this problem and
found that maximum entropy principle applied to the Gibbs--Shannon
entropy could give rise to the power--law distribution under only
very special constraint that "has not been considered as a natural
one for use in auxiliary conditions."

\section{Helmholtz free energy and Renyi entropy}

The well--known Boltzmann formula \index{Boltzmann entropy}
defines a statistical entropy, as a logarithm of a number of
states $W$ attainable for the system
\begin{equation}
S_W^{(B)}=\ln W
\end{equation}
Here and below the entropy is written as dimensionless value
without the Boltzmann constant $k_B$. Besides, we will use a
natural logarithm instead of binary logarithm accepted in
information theory.

This definition is valid not only for physical systems but for
much more wide class of social, biological, communication and
other systems described with the use of statistical approach. The
only but decisive restriction on the validity of this equation is
the condition that all $W$ states of the system have equal
probabilities (such systems are described in statistical physics
by a microcanonical ensemble). It means that probabilities
$p_i=p_W\equiv 1/W$ (for all $i =1, 2, ..., W$) that permits to
rewrite the Boltzmann formula as $ S_W^{(B)}=- \ln p_W.$ When the
probabilities $p_i$ are not all equal we can introduce an ensemble
of microcanonical subsystems in such a manner that all $W_i$
states of the $i$-th subsystem have equal probabilities $p_i$ and
its Boltzmann entropy is $S_i^{(B)}=- \ln p_i$. The simple
averaging of the Boltzmann entropy $S_i^{(B)}$ leads to the
Gibbs--Shannon entropy \index{Gibbs--Shannon entropy}
\begin{equation} \label{3}
S^{(G)}=\langle S_i^{(B)}\rangle_p \equiv - \sum_i p_i\ln p_i.
\end{equation}
Just such derivation of $S^{(G)}$ is used in some textbooks (see,
e. g. \cite{Haken,Nicolis})

This entropy is generally accepted in equilibrium and
non-equilibrium statistical thermodynamics \cite{Zub1,Zub2} and
communication theory but needs in modification for complex
systems. To seek out a direction of modification of the
Gibbs--Shannon entropy we consider first extremal properties of an
equilibrium state in thermodynamics.

A direct calculation of an average energy of a system gives the
internal energy  $U$, its extremum is characteristic of an
equilibrium state of rest for a mechanical system, other than a
thermodynamic system that can change heat with a heat bath of a
fixed temperature $T_0$. An equilibrium state of the latter system
is characterized by extremum of the Helmholtz free energy $F$. To
derive it statistically from the Hamiltonian $H = \sum_i H_i$
without use of thermodynamics we can following Balescu
\cite{Balescu} introduce a moment-generating function
\begin{equation}
\Phi_H (\alpha)=\sum_i e^{ \alpha H_i},
\end{equation}
where $\alpha$ is the arbitrary constant, and construct a
cumulant-generating function
\begin{equation}
\Psi_H (\alpha)=\ln \Phi_H (\alpha)
\end{equation}
that becomes the Helmholtz free energy $F$ when divided by
$\alpha$ that is chosen as $\alpha ~ =~ -1/k_B T_0$.

Such a choice of the pre-factor $1/\alpha$ ensures a limiting
passing of the Helmholtz free energy $F$ \index{Helmholtz free
energy} into the internal energy $U$  when $\alpha\to\infty$
($T_0\to 0$).

Now we return to the problem of a generalized entropy for open
complex systems. Exchange by both energy and entropy (or
information) with the environment is inherent in such systems (see
e. g. the book \cite{Haken} devoted to this subject).

It is pertinent to introduce the noun of an {\it entropy bath} (or
{\it information bath}). Coupling with the entropy bath can be
regarded as a necessary condition for self-organization of a
complex system.

As a result of such coupling the system under consideration can
not reach a state of thermodynamic equilibrium that is
characterized by maximum of the Gibbs--Shannon entropy, derived by
the simple averaging of the Boltzmann entropy. It is necessary to
look for any other function to characterize its steady state
resulted from the coupling with the entropy bath.

An effort may be made to find a "free entropy" of a sort by the
same way that was used above for derivation of the Helmholtz free
energy for a system coupled with a heat bath. The
moment-generating function is introduced as
\begin{equation} \label{6}
\Phi_S (\alpha)=\sum_i e^{ \alpha S_i^{(B)}}
\end{equation}
Then the cumulant-generating function is
\begin{equation}
\Psi_S (\alpha)=\ln \Phi_S (\alpha)=\ln\sum_i p_i^{- \alpha}.
\end{equation}
To obtain the desired generalization of the entropy we are to find
an $\alpha$-dependent numerical pre-factor which ensures a
limiting pass of the new entropy into the Gibbs--Shannon entropy.
Such the coefficient is  $1/(1+\alpha)$. Indeed, the new
$\alpha$-family of entropies
\begin{equation}
S(\alpha)=\frac 1{1+\alpha}\ln\sum_i p_i^{- \alpha}.
\end{equation}
includes the Gibbs--Shannon entropy as a particular case when $
\alpha\to -1$.

Thus, it has appeared that the desired "free entropy" coincides
with the known Renyi entropy  \cite{Renyi}. It is conventional to
write it with the parameter $q = -\alpha$ in the form
\begin{equation} \label{9}
S_q^{(R)}(p)=\frac 1{1-q}\ln\sum_i p_i^{q}
\end{equation}

On the other hand, we can represent Eq. (\ref{6}) as
\begin{equation}
\Phi_S (q)=\sum_i p_i e^{(1- q)S_i^{(B)}}
\end{equation}
Then the Renyi entropy can be represented as a particular case of
the Kolmogorov--Nagumo \cite{Kolm,Nagumo} generalized averages
\begin{equation}
\langle x \rangle_\phi=\phi^{-1}\left(\sum_i p_i\phi (x_i)\right )
\end{equation}
if we put there the Kolmogorov--Nagumo function in the form $\phi
(x)= \exp\{(1-q)x\}$, $\phi^{-1} (x)=\ln x^{1/(1-q)}$ and choose
$x_i=\ln S_i^{(B)}$.

Renyi introduced  his entropy just in such a manner. Renyi wanted
to find the most general class of entropies which preserved the
additivity for statistically independent systems and was
compatible with the Kolmogorov--Nagumo generalized average. By
this way he found the exponential Kolmogorov--Nagumo function
$\phi (x)$. Physically, such a choice of $\phi (x)$ on its own
appears accidental until it is not pointed to the fact that the
same exponential function of the Hamiltonian provides derivation
of the free energy which is extremal in an equilibrium state of a
thermodynamic system exchanging heat with a heat bath. This fact
permits us to suppose that the Renyi entropy derived in the same
manner should be extremal at a steady state of a complex system
which exchange entropy with its surroundings actively.

Note that for linear $\phi (x) = c\,x + d $ the Kolmogorov--Nagumo
generalized average turns out to be the ordinary linear mean and
hence the Gibbs-Shannon entropy follows as an average entropy in
the usual sense.

\section{Axiomatical foundation of the Renyi entropy}

In view of the way of the Renyi entropy derivation we can suppose
that it is maximal at a steady state of a complex system being in
contact with the entropy bath. Such a supposing is justified by
the Shore--Johnson theorem \cite{Shore,Shore2,Uffink}. They
considered a procedure of updating of a distribution function when
a new information related to the system had appeared in a form of
an additional constraint $I$.

Shore and Johnson gave five ``consistency axioms'' for this
updating operation \cite{Shore,Shore2}: 1) Uniqueness: The result
should be unique. 2) Invariance: The choice of coordinate system
should not matter (for continuous probability densities).3) System
independence: It should not matter whether one accounts for
independent information about independent systems separately in
terms of different densities or in terms of a joint density. 4)
Subset independence: It should not matter whether one treats
disjoint subsets of system states in terms of separate conditional
densities or in terms of the full density. 5) In the absence of
new information, we should not change the prior.

The following theorem is proven on the base of these axioms (here
it is for the particular case of a homogeneous prior distribution
$u_i=1/W$ (for all $i$)):\\
{\bf Theorem}
{\it An updating procedure satisfies the five consistency axioms
above if and only if it is equivalent to the rule\\
Maximize
$$U_{\eta}(p)=(\sum_i^Wp_i^{1-\eta})^{1/\eta},\,\,(\eta<1)$$ under
the constraint
$I$.}\\
or\\
{\it Maximize any monotonous function $\Psi (U_{\eta}(p))$ under
the constraint $I$.}

The most evident choice of the monotonous function is $\Psi
(U_\eta)= \ln\,U_{\eta}(p)$, that is the Renyi entropy $S_q^{(R)}$
for $q=1-\eta$. Such a choice of $\Psi$ ensures the limit
$S_q^{(R)}\to S^{(G)}$ when $q\to 1$ and passage of $S_q^{(R)}$
(for all $q$) to $S_W^{(B)}$ in the case of absence of new
information when $p_i=u_i=1/W$ (for all $i$). Both these
properties should be considered as necessary conditions for a
choice of $\Psi (U_\eta)$. In particular, the function $\Psi_T
(U_\eta)= (U^\eta_{\eta}-1)/\eta$ leading to the Tsallis entropy
fails because it does not satisfy the second of these conditions.

Thus, the Shore-Johnson theorem provide quite conclusive
foundation of the Renyi entropy as itself and the maximum entropy
principle for it and in doing so it justifies the above proposal
that the Renyi entropy as the free entropy is maximal at a steady
state of a complex system.

In the light of this theorem the Khinchin's uniqueness theorem
\cite{Khinchin} for the Gibbs--Shannon entropy should be
reconsidered.  Khinchin based on the next three axioms:\\
(1) $S(p)$ is a function of the probabilities $p_i$ only and has
to take its maximum value for the uniform distribution of
probabilities $p_i = 1/W$: $ S(1/W,...,1/W)\geq S(p')$, where $p'$
is any other distribution.\\
(2) The second axiom refers to a composition $\Sigma$ of a master
subsystem $\Sigma^I$ and subordinate subsystem $\Sigma^{II}$ for
which probability of a composed state is
\begin{equation} \label{4.1}
p_{ij} = Q(j|i)p^I_i
\end{equation}
where $Q(j|i)$ is the conditional probability to find the
subsystem $\Sigma^{II}$ in the state $j$ if the master  subsystem
$\Sigma^{I}$ is in the state $i$. Then the axiom requires that
\begin{equation} \label{4.2}
S(p) = S(p^I) + S(p^I|p^{II})
\end{equation}
where
\begin{equation} \label{4.3}
S(p^I|p^{II})=\sum_i p^I_i S(Q|i)
\end{equation}
is the conditional entropy and $S(Q|i)$ is the partial conditional
entropy of the subsystem  $\Sigma^{II}$ when the subsystem
$\Sigma^{I}$ is in the $i$-th state.\\
(3) $S(p)$ remains unchanged if the sample set is enlarged by a
new, impossible event with zero probability: $S(p_1,...,\,p_W) =
S(p_1,...,\,p_W,\,0)$

While proving his uniqueness theorem Khinchin enlarged the second
axiom. He supposed that all $U$  states of the composite system
$\Sigma$ were equally probable, that is, $p_{ij}=1/{U}$ for all
$i$ and $j$; whence he got
\begin{equation} \label{4.4}
S(p) = \ln U
\end{equation}
Besides, Khinchin supposed that $U^{II}_i$  states of the
subsystem $\Sigma^{II}$ corresponding to each $i$-th state of the
master subsystem $\Sigma^I$ are equally probable, as well. So, he
took $Q(j|i)=1/{U^{II}_i}$ for $i=1,...,W$ and $j\in U_i ^{II}$;
whence from Eqs. (\ref{4.1}) and (\ref{4.3}) he obtained
$U^{II}_i= U p^I_i$ and
\begin{equation} \label{4.5}
 S(Q|i)= \ln {U^{II}_i},\,\,\,S(p^I|p^{II})=\ln
{U}+\sum_i p^I_i \ln p^I_i.
\end{equation}
Substituting Eqs. (\ref{4.4}), (\ref{4.5}) into Eq. (\ref{4.2})
Khinchin got the Gibbs--Shannon entropy for the master subsystem
$\Sigma^I$ $S^{(G)}(p^I)=-\sum_i^W p^I_i\ln p^I_i$.
It should be noted that the last Khinchin's supposition of equal
probabilities of states of the subsystem $\Sigma^{II}$ is the most
questionable and should be abandoned. Indeed, the probability
distribution for $\Sigma^{II}$ coupled with the master subsystem
should be rather canonical distribution than equally probable one.
On the other hand, the abandonment of this supposition destroys
all the proof of the Khinchin theorem.

So, we are left with only the Shore--Johnson theorem where the
Gibbs--Shannon entropy is no more than a particular case of the
Renyi entropy.
\section{The Renyi thermostatistics}

The maximum entropy principle for the Renyi entropy $S^{(R)}$
under additional constraints of fixed value $U=\langle
H\rangle_p\equiv\sum_i H_i p _i$, and normalization of $p$ gives
rise \cite{Bash,Bash1} to the Renyi distribution function
\begin{eqnarray}\label{RD}
p_i&=& p_i^{(R)} =Z_R^{-1}\left(1-\beta\frac{q-1}{q}\Delta
H_i\right)^{\frac{1}{q-1}}\\
Z_R&=&\sum_i\left(1-\beta\frac{q-1}{q}\Delta
H_i\right)^{\frac{1}{q-1}},\,\,\Delta H_i= H_i-U.\nonumber
\end{eqnarray}
At $q\to 1$ the distribution $\{p^{(R)}_i\}$ becomes the Gibbs
canonical distribution in which the constant $\beta=1/k_BT_0$.

Substituting the Renyi distribution (\ref{RD}) into the Renyi
entropy definition (\ref{9}), we find the thermodynamic entropy in
the Renyi thermostatistics as
\begin{equation}
\tilde S_q^{(R)}=S^{(R)}_q (p^{(R)}_q) =k_B\ln Z_q^{(R)}.
\end{equation}
where the Boltzmann constant $k_B$ is introduced.

When $q\to 1$ this entropy passes into thermodynamic entropy in
the Gibbs thermostatistics
\begin{equation}
\tilde S^{(G)}=S^{(G)}(p^{(G)}) = k_B\ln \sum^W_i e^{-\beta\Delta
H_i}.
\end{equation}
On the other hand, the Gibbs thermostatistics is based on the
Gibbs distribution for energy $H_i$ but not fluctuations of energy
$\Delta H_i$. So, the Renyi distribution (\ref{RD}) should be
represented as a distribution for energy $H_i$, as well. Dividing
both the expression in the brackets in Eq. (\ref{RD}) and $Z_R$ by
$(1-(1-q)\beta U/q)^{1/(q-1)}$ we get the alternative equivalent
form of the Renyi distribution
\begin{equation}\label{RD*}
p_i^{*} =Z_*^{-1}\left(1+\frac{1-q}{q}\beta^*
H_i\right)^{\frac{1}{q-1}}
\end{equation}
$$
Z_*=\sum_i\left(1+\frac{1-q}{q}\beta^*
H_i\right)^{\frac{1}{q-1}},\,\,\beta^*=\frac{\beta}{1-\frac{1-q}{q}\beta
U} $$

The thermodynamic Renyi entropy can be represented in the
alternative form, as well
\begin{eqnarray}\label{RE*}
\tilde S_q^{(R)}(p^{*})&=&\frac {k_B}{1-q}\ln\sum_i p_i^{*q}\nonumber\\
&=&\frac {k_B}{1-q}\ln (1+\frac{1-q}{q}\beta^* U)+k_B\ln Z_*
\end{eqnarray}

A physical $q$-dependent temperature in the Renyi superstatistics
should be defined in a standard manner as
\begin{equation}
T_q=\left(\frac{\partial S_q^{(R)}}{\partial
U}\right)^{-1}=\left(\frac{k_B}{q}\frac {\beta^*}
{1+\frac{1-q}{q}\beta^* U}\right)^{-1}=\frac {q}{k_B\beta}.
\end{equation}
It was shown \cite{Bash} that $\beta=\beta_0$ at least for the
power-law Hamiltonian, so the physical $q$-dependent temperature
becomes 
$T_q=q T_0$
where $T_0$ is the temperature of a heat bath. According to Ref.
\cite{Klim}, the fact that $T_q < T$ $(q<1)$ says in favor of
greater ordering of states with lower $q$.

The Helmholtz free energy is defined in the Renyi thermostatistics
as
\begin{equation}
F^{(R)}= -k_BT_q\ln Z_*.
\end{equation}
On the other hand, the thermodynamic definition of the free energy
should be
\begin{equation}
\tilde F= U -T_q\tilde S^{(R)}.
\end{equation}
It is not difficult to ensure that both definitions coinside in
the limit $q\to 1$. For arbitrary $q$, their equivalence can be
checked for the particular case of the power-law Hamiltonian. With
the use of the relation $\beta U=1/\kappa$ \cite{Bash} we get
$\tilde F -F^{(R)}=C_qT_q$ where
\begin{equation}
C_q= 1/(q\kappa)-\ln[1-(1-q)/(q\kappa +q-1)]/(1-q).
\end{equation}
Such a difference can not provoke objections because of the free
energy is determined in thermodynamics with an accuracy of an
addent $C_1T+C_2$.

\section{Entropic phase transition}
Thus, we have Gibbs and Renyi thermostatistics based
on different microscopic entropy definitions. Each of them
provides an adequate description of corresponding class of systems
and we need in a rigorous formulation of conditions of transfer
from one thermostatistics to another.

Transfer from the Gibbs distribution describing a state of dynamic
chaos \cite{Klim} to power--law Renyi distributions that are
characteristic for ordered self-organized systems \cite{Bak}
corresponds to an increase of an "order parameter" $\eta =1-q$
from zero at $q=1$ up to $\eta_{max} =1-q_{min}$.

In accordance to the Landau theory \cite{Landau} of phase
transitions  an entropy derivative with respect to the order
parameter undergoes a jump at a point of the phase transition.

Here we deal with the transfer from the Gibbs thermostatistics to
the Renyi thermostatistics corresponding to non-zeroth values of
the order parameter $\eta $. Let us consider a variation of the
entropy at this transition.

Now it is not difficult to calculate the limiting value at $\eta
\to 0$ of the derivative of the entropy $\tilde S_\eta^{(R)}$ with
respect to $\eta $. We get
\begin{equation}
\lim_{\eta\to 0}\left(\frac{d\tilde S^{(R)}}{d\eta}
\right)=\frac{k_B}{2}\beta^2 \sum^W_i p^{(G)}_i\left(\Delta
H_i\right)^2
\end{equation}
According to a fluctuation theory for the Gibbs equilibrium
ensemble we have
\begin{equation} \label{54}
\sum^W_i p^{(G)}_i\left(\Delta H_i\right)^2=
\frac{1}{k_B\beta^2}\frac{dU}{dT}=\frac{1}{k_B\beta^2}C_V
\end{equation}
whence
\begin{equation} \label{55}
\lim_{\eta\to 0}\left(\frac{d\Delta\tilde S}{d\eta}
\right)=\frac{1}{2}C_V.
\end{equation}
where $C_V$ is the heat capacity at a constant volume.

Thus, the derivative of the entropy gain with respect to the order
parameter exhibits the jump (equal to $C_V/2$) at $\eta =0$. This
permits us to consider the transfer to the Renyi thermostatistics
as a peculiar kind of a phase transition into a more organized
state. We can give this transition the name {\it entropic phase
transition}.

As a result of the entropic phase transition the system passes
into an ordered state with the order parameter $\eta \neq 0$. In
contrast to the usual phase transition that takes place at the
temperature of phase transition, conditions of the entropic phase
transition are likely to be determined partially for each concrete
system. For example, a threshold of emergence of turbulence (see
\cite{ZubMor}) as an ordered structure is determined by a critical
Reynolds number and an emergence of Benard cells is determined by
a critical Rayleigh number (see \cite{Klim}).

Social, economical and biological systems are realized as a rule
in ordered self--organized forms. This is the reason why
power--law and closely related distributions are characteristic
for them but not canonical Gibbs distribution.

\begin{figure}[t]
\begin{minipage}{.50\linewidth}
 \centering\epsfig{figure=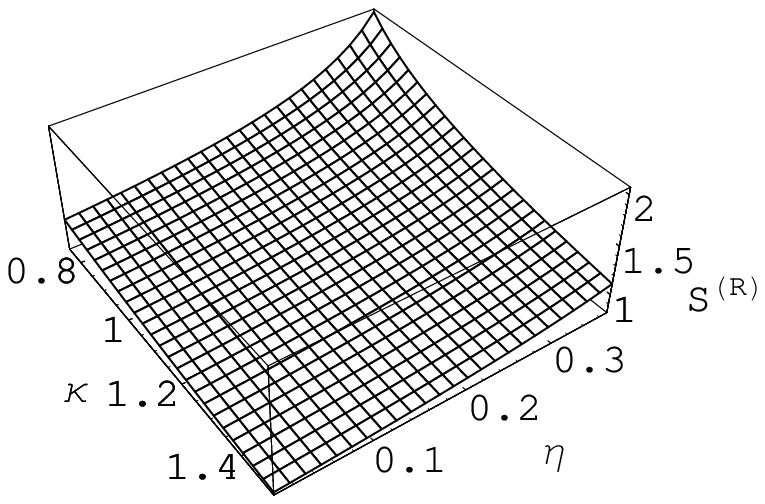, height=5.6cm}
 \end{minipage}
\begin{minipage}{.52\linewidth}
 \centering\epsfig{figure=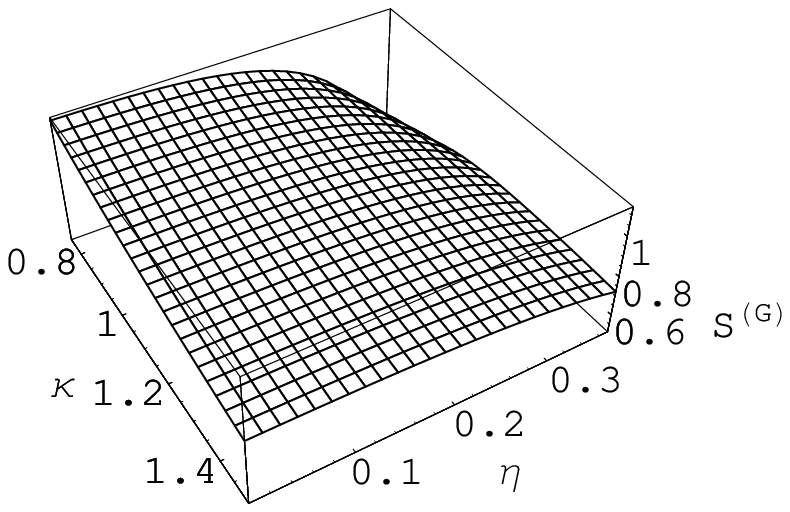, height=5.6cm}
  \end{minipage}
\caption{The landscapes of the entropies $\tilde
S_\eta^{(R)}[p^{(R)}(\eta,\kappa)]$ (left) and $\tilde
S^{(G)}[p^{(R)}(\eta,\kappa)]$ (right) for the power--law
Hamiltonian with the exponent $\kappa$ and
$\eta<\kappa/(1+\kappa)$.}
\end{figure}

For the particular case of a power--law Hamiltonian
$H_i=Cx_i^\kappa$ problem of a value of the order parameter was
discussed in Ref. \cite{Bash} where maximum maximorum of the
thermodynamic Renyi entropy was found at $q=q_{min}=1/(1+\kappa)$,
that is, at $\eta=\eta_{max}=\kappa /(1+\kappa)$. The  Renyi
distribution for such $\eta_{max}$ becomes a pure power--law
distribution that agrees with observable data for self-organized
systems.

The landscapes of this entropy $\tilde
S_\eta^{(R)}[p^{(R)}(\eta,\kappa)]$ is illustrated in Fig. 1
(left). The landscape of the usual thermodynamic Gibbs entropy
$S^{(G)}[p^R(x|q,\kappa)]$ for the same Renyi distribution is
illustrated in Fig. 1 (right). It is seen that in contrast to
$\tilde S_\eta^{(R)}[p^{(R)}(\eta,\kappa)]$ the Gibbs entropy
$S^{(G)}[p^R(x|q,\kappa)]$ decreases with the gain of $\eta$ and
attains its maximum at $\eta=0$, that is in the most disordered
state when the Renyi distribution becomes the Gibbs canonical one.

\section{Conclusion}
It should be stressed that thermodynamic laws are irrelevant to
microscopic interpretations of thermodynamic functions. On the
other hand, according to the Boltzmann--Gibbs microscopic
interpretation of entropy, its gain is accompanied by evolution of
a system to an homogeneous equilibrium state of thermal chaos. In
contrast, the Renyi thermodynamic entropy increases as a system
ordering (departure of the order parameter $\eta$ from zero)
increases (see Fig. 1). So, it can be considered as a kind of
potential that drives the system to self-organized state.

Transfer from the usual Gibbs thermostatistics to the Renyi
thermostatistics takes the form of a phase transition of ordering
with the order parameter $\eta$. As soon as the system passes into
this new phase state of the Renyi thermostatistics, a spontaneous
development of self--organization to a more ordered state begins
accompanied with gain of thermodynamic entropy. In doing so the
well-known contradiction between observable spontaneous
self--organization and the Second Law is eliminated when we use
the Renyi entropy as a microscopic definition of the thermodynamic
entropy instead of the Gibbs--Shannon one.

Moreover, it may be supposed that biological evolution or
development are governed by the extremal principle of the Renyi
thermostatistics.\\
\noindent {\bf Acknowledgement.} It is pleasure to thank A. V.
Vityazev for many fruitful discussions and supporting this work. I
must to pay homage to late Prof. D. N. Zubarev, my teacher who
payed my attention to the Renyi entropy many years ago.

\end{document}